\documentclass[preprint,3p]{elsarticle}

\usepackage{graphics}
\usepackage{graphicx}
\usepackage{multirow}
\usepackage{booktabs}
\usepackage{epsfig}
\usepackage{amssymb}
\usepackage{amsthm}
\usepackage{color}
\usepackage{natbib}
\usepackage{url}
\usepackage{subfigure}
\graphicspath{{figures/}}

\usepackage{lineno}
\journal{Nucl. Instr. and Meth. A}
\begin{document}
\begin{frontmatter}
%%%%%%%%%%%%%%%%%%%%%%%%%%%%%%%%%%%%%%%%%%%%%%%%%%%%%%%%%%%%%%%%%%%%
\title{Scenarios for LHC/FCC Based Gamma-Proton Colliders}
\author[nigde]{Zafer Nergiz\corref{bilm}}
\cortext[bilm]{Corresponding Author tel: +903882254080}
\ead{znergiz@ohu.edu.tr}
\author[TOBB,ANAS]{Saleh Sultansoy}
\author[KSU]{Husnu Aksakal}
\author[CERN]{Frank Zimmermann}
\address[nigde]{Nigde Omer Halisdemir University, Faculty of Arts and Sciences, Department of Physics, 51200 Nigde, Turkey}
\address[TOBB]{TOBB University, Ankara, Turkey}
\address[ANAS]{Institute of Physics, Academy of Sciences, Baku, Azerbaijan}
\address[KSU]{Kahramanmaras Sutcu Imam University, Kahramanmaras, Turkey}
\address[CERN]{CERN, Switzerland}

%%%%%%%%%%%%%
\begin{abstract}

The advantage of the linac-ring type electron proton collider is that it allows for the straightforward construction of  $\gamma$p collider. In a  $\gamma$p collider high energy photons can be generated from Compton backscattering of laser photons off electrons from a linear accelerator. In this study main parameters of photon-proton colliders based on some future electron linear accelerator projects and protons supplied form LHC or FCC are evaluated.

\end{abstract}
\end{frontmatter}

%%%\linenumbers

\section{\label{intro}Introduction} 

Linac-Ring type ep colliders seems to be sole realistic way to handle Multi-TeV center of mass energy in electron
proton collisions \cite{akay10}. Today 60 GeV energy recovery linac is considered as baseline option 
both for LHC and FCC based ep colliders \cite{ bordry18}. On the other hand energy frontier options 
using one pass linacs have a huge potential for BSM physics search \cite{sultansoy17}.

Combination of several future linear accelerator projects with LHC (Large Hadron Collider) and FCC (Future Circular Collider) 
offers a unique opportunity to build $\gamma$p colliders \cite{sultansoy17,  aksakal07, nergiz07}.  $\gamma$p collisions 
allow investigations of extremely low x and high $Q^{2}$ physics in quantum chromodynamics. 
Physics search potential of $\gamma$p colliders 
at a new kinematic range is reviewed in references \cite{sultansoy98, ciftci95, engin03, sultansoy04}. 

\begin{figure}[htbp]
\centering
\includegraphics[width=0.55\columnwidth]{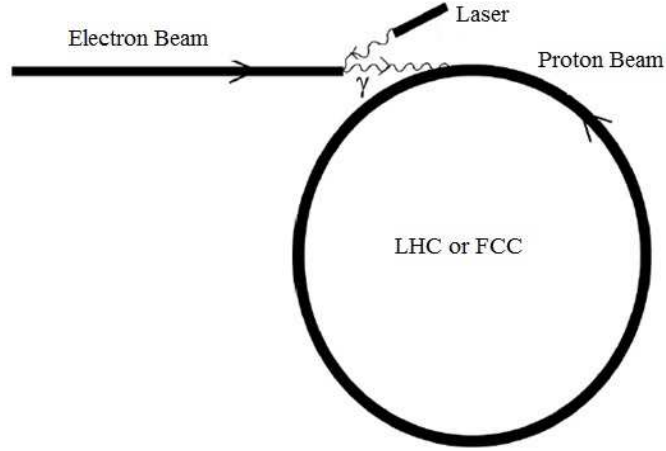}
\caption {Schematic view of  $\gamma$p colliders.}
\label{fig:layout}
\end{figure}

In the photon colliders ($\gamma \gamma$,  $\gamma$e,  $\gamma$p and  $\gamma$A), high energy photons are produced by the Compton backscattering 
of the intense laser pulse off the electron beam
provided by the linear accelerator \cite{telnov90, telnov95, borden}. In our case, the backscattered photons are generated at conversion point (CP) and are 
collided with protons at interaction point (IP). A schematic view of a $\gamma$p collider is shown in Fig.~\ref{fig:layout}. 

\section{\label{sec:compton} Compton Backscattering}

Compton cross section is characterized by a dimensionless parameter given by~\cite{ginzburg83}
\begin{equation}
x\;=\;\frac{4E_{b}\omega_{0}}{m_{e}^{2}}Cos^{2}\left(\frac{\alpha_{0}}{2}\right)\;,
\label{eq:xparam}
\end{equation}
where $E_{b}$ is the initial energy of electrons, $\omega_{0}$ is the energy of a 
laser photon and $\alpha_{0}$ is the collision angle between laser beam and electron beam. In the case of head on collision and with practical 
units,  Eq.~(\ref{eq:xparam}) can be written as $x=15.3E_{b}[TeV]\omega_{0}[eV]$. The energy of backscattered photons increases with increasing value of 
the parameter $x$. However, if $x$ is larger than 4.8, high energy photons can be lost due to $e^{+}e^{-}$ pair 
creation in collisions of backscattered photons with unscattered  laser photons. Maximum energy of 
backscattered photons is $\omega_{max}=E_{b}x/(x+1)$. For x=4.8 and neglecting nonlinear Compton scattering process, maximum photon energy is 0.83$\times E_{b}$. 

As the laser photon density increases, the collision probability of two or more laser photons with a high energy electron is increases as well. Therefore, 
in the strong electromagnetic fields at the laser focus, nonlinear Compton scattering process given by  
\begin{equation}
 e^{-}+n\gamma_{laser}\rightarrow e^{-}+\gamma \;,  \;\;\;\;   (n\geq 1)    
\end{equation}
becomes important. This nonlinear effect is characterized by the parameter
\begin{equation}
 \xi^{2}=n_{\gamma}\left(\frac{4\pi\alpha}{m_{e}^{2}\omega_{0}}\right), 
\end{equation}
where $n_{\gamma}$ is the laser photon density, $\alpha$ is fine structure constant and $m_{e}$ is electron mass. Considering nonlinear Compton scattering process, maximum energy of backscattered 
photons produced by electron colliding with n laser photons is given by~\cite{galynskii01}
\begin{equation}
\omega_{max}^{n}=E_{b}\frac{nx}{1+\xi^{2}+nx}.
\end{equation} 

Another process, which affects the backscattered photon spectrum is successive scattering. To obtain the effects of the successive scattering,
simulation program is needed.

\section {\label{sec:lumi} Luminosity of LHC/FCC Based $\gamma$p Colliders}

Two hadron colliders (LHC and FCC) with several lepton collider projects (ILC, CLIC and PWFA-LC) as well as ERL 60  offer the possibility to realize $\gamma p$ colliders in different range  of energy. The energy spectra of backscattered photons from different linear accelerator projects and luminosity spectra of different linac$\times$LHC/FCC based $\gamma p$ colliders are investigated by using CAIN simulation code~\cite{cain}.  

\subsection {\label{sec:linac} Linear collider projects}

Because of the high energy losses of electrons due to the synchrotron radiation, two linear accelerator 
projects for future lepton colliders have priority: ILC (International
Linear Collider) and CLIC (Compact Linear Collider). 

The ILC  is designed as a 500 GeV 
center-of-mass energy linear electron-positron collider based on superconducting radio-frequency technology and can be extended to 1 TeV \cite{Adolphsen}. 

CLIC is a future collider project to provide $e^{+}e^{-}$ collisions with normal conducting high frequency (12 G Hz) rf structure. The CLIC is planned to construct 
in three stage: 0.380 TeV, 1.5 TeV and 3 TeV center of mass energy, respectively  \cite{bordry18}. Besides this three stages, 60 GeV option was also proposed 
in ref. \cite{schulte04}. Recently, using multiple delay loops and rf deflectors have been proposed to match the bunch structure of proton beam and CLIC 60 GeV \cite{frank18}.   
The electron beam will has 312 bunches with 25 ns spacing and 25 beam pulses spaced 6 $\mu s$  repeating at 100 Hz. Therefore, the collision frequency of 
312$\times$25$\times$100 can be achieved.
 In this paper upgraded version of 60 GeV option is choosen for CLIC as electron source. 

LHeC (Large Hadron electron Collider) is a project which aim to interact the 60 GeV electrons with LHC's protons. As electron source,
 Energy Recovery Linac (ERL) is proposed to provide high electron average current and increase the
luminosity  \cite{lhec11}. The ERL can also be used for e source for compton backscattering process. However, because of the energy losses of 65$\%$  of electrons after 
conversion region the Energy Recovery Process should be bypassed. In that case the average electron beam current of the electron linac can be 0.3 mA.    

Recently, more compact linear collider based beam driven plasma wake field technology is proposed. In PWFA-LC proposal, extremely high electron beam energy of 5 TeV
 can be realized with relatively low cost and high efficiency  \cite{delahaye14}.   

The mentioned  linac parameters are shown in Tables~\ref{elhcparam} and \ref{efccparam}.

\subsection {\label{sec:laser} Laser requirements}

If the multiple scattering is neglected and assuming that the laser profile seen by each electron is the same, 
the conversion probability of generating high energy gamma photons per individual electron can be written as  \cite{nlc}

\begin{equation}
p=1-e^{-q},
\label{eq:pparam}
\end{equation}
where
\begin{equation}
q=\frac{\sigma_{c}A}{\omega_{0}\Sigma_{L}},
\label{eq:qparam}
\end{equation}
where A is laser pulse energy, $\omega_{0}$ is  laser photon energy, $\Sigma_{L}$ is the transverse area of laser spot and
$\sigma_{c}$ is the Compton cross section which is $1.75 \times 10^{-25} cm^2$ for x=4.8. 
Proposed laser parameters are given in Table~\ref{lparam}

\begin{table}[htbp]
    \caption{Laser parameters for the 60 GeV ERL Options and CLIC, 500 GeV ILC and 5000 GeV PWFA-LC.   }\label{lparam}
\smallskip
\centering
\small\addtolength{\tabcolsep}{-2pt}
    \begin{tabular}{lccc}\hline
  %% Parameters                          & 60 GeV ERL Options  & 500 GeV ILC  & 5000 GeV PWFA \\\hline
 Parameters                          & ERL and CLIC  & ILC  & PWFA-LC \\\hline
  Laser wavelength ($\mu m$)   & 0.24                          & 2.0                 & 20  \\
    Pulse Energy (J)                  & 10                             & 2                     & 8  \\
   Pulse length  (mm)                & 1                                & 0.3366           & 2.5	    \\
  Rayleigh length  (mm)            & 0.5                             & 0.5                  & 3.5  \\
  $\xi^{2}$                            & 0.05                             & 0.23           &  0.16   \\\hline    
   \end{tabular}
%    \end{ruledtabular}
\end{table}

\begin{figure}[htbp]
\centering
\includegraphics[width=0.55\columnwidth]{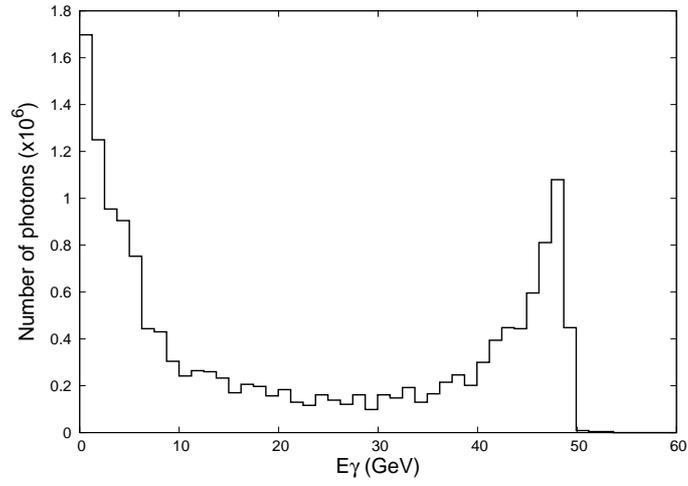}
\caption {Backscattered photon spectrum from LHeC-ERL Linac.}
\label{fig:lhecgspec}
\end{figure}

\begin{figure}[htbp]
\centering
\includegraphics[width=0.55\columnwidth]{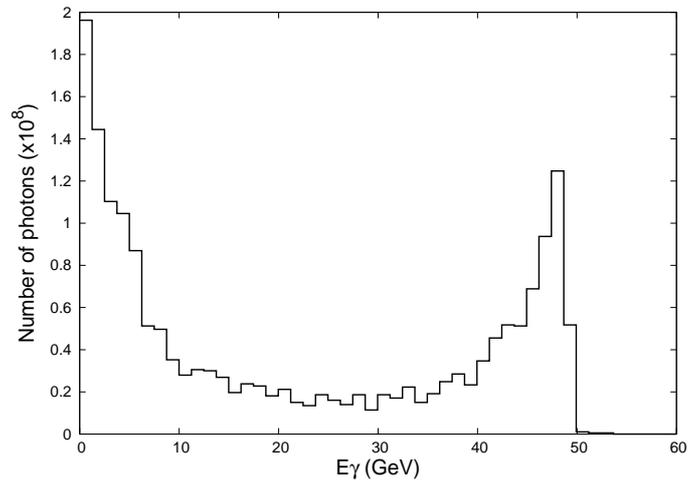}
\caption {Backscattered photon spectrum from CLIC 60 GeV Linac.}
\label{fig:clicgspec}
\end{figure}

\begin{figure}[htbp]
\centering
\includegraphics[width=0.55\columnwidth]{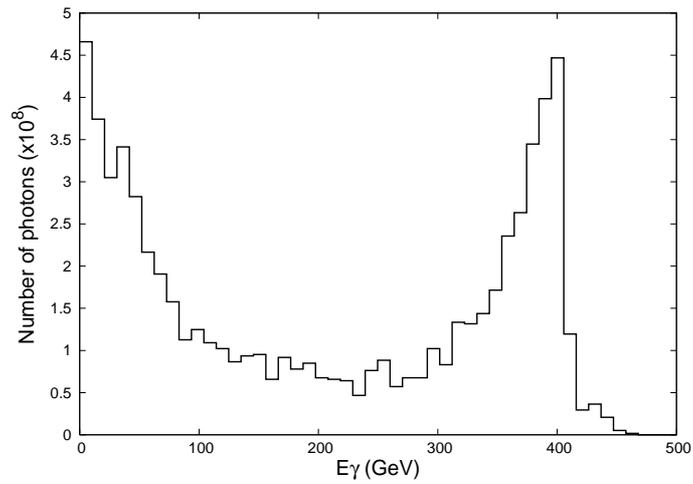}
\caption {Backscattered photon spectrum from ILC Linac.}
\label{fig:ILCgspec}
\end{figure}

\begin{figure}[htbp]
\centering
\includegraphics[width=0.55\columnwidth]{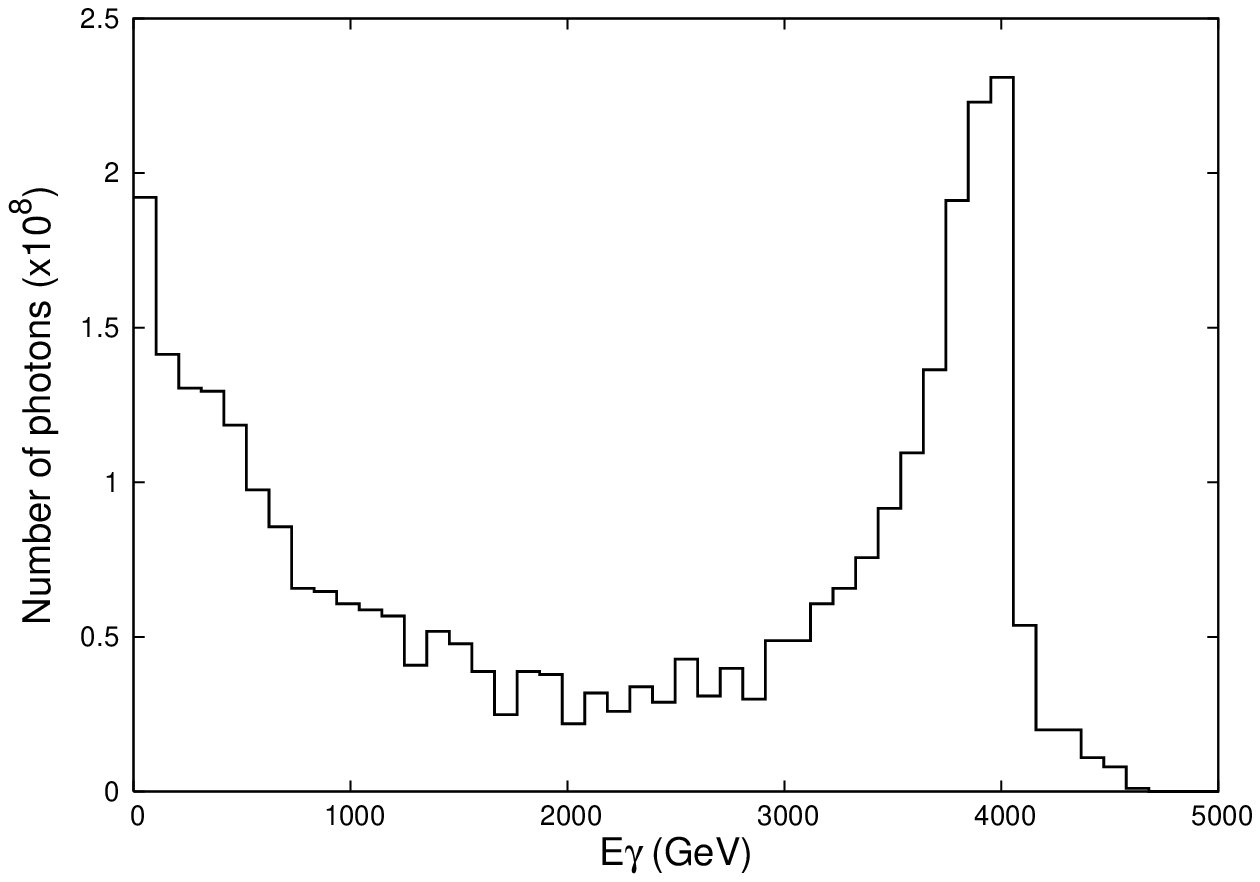}
\caption {Backscattered photon spectrum from PWFA-LC Linac.}
\label{fig:PWFAgspec}
\end{figure}

Backscattered photon spectra obtained from CAIN simulation code are shown in  Figs.~\ref{fig:lhecgspec},~\ref{fig:clicgspec},  ~\ref{fig:ILCgspec} and~\ref{fig:PWFAgspec}. 
A peak seen at low energies is a result of the successive scattering. The spreads at high energy regions in the spectra are due to the nonlinear Compton backscattering effect.

\subsection {\label{sec:glhc} LHC based $\gamma$p colliders}
While the proton-proton collisions is running at LHC (Large Hadron Collider), the design studies on  several post-LHC hadron collider projects are studied at CERN: 
HE-LHC (High Energy LHC), HL-LHC (High Luminosity LHC) and FCC (Future Circular Collider) options \cite{bordry18}. The High-Luminosity LHC is an approved luminosity upgrade of the LHC.
The High Energy LHC aimed to reach the beam energy around 13.5 TeV in the existing LHC tunnel by using 16.0 Tesla FCC dipole magnets instead of LHC's 8.33 Tesla nominal dipol magnets.
The proposed linacs and LHC options beam parameters are given in Table~\ref{elhcparam}. The parameters for LHeC CDR,  HL-LHC and HE-LHC parameters updated from ref. \cite{bordry18} according to the $\gamma$p colliders requirements. 

\begin{table*}[htbp]
    \caption{Proposed accelerator parameters for different e-LHC based $\gamma$p colliders. }\label{elhcparam}
\smallskip
\centering
\small\addtolength{\tabcolsep}{-2pt}
    \begin{tabular}{lcccccc}\hline
    Parameters                          & LHeC CDR	  & ep at HL-LHC  & ep at HE-LHC   &CLIC-LHC    & ILC-LHC   & PWFA-LHC \\\hline
  $E_{p}$  (TeV)                      & 7                        & 7                     & 12.5                 &  7           & 7                 &7  \\
    $E_{e}$  (GeV)                   & 60                      & 60                   & 60                     &  60          & 500             & 5000	    \\
   Max. C.M.E (TeV)                          & 1.17          & 1.17               & 1.6          	   & 1.17       & 3.40          & 10.75	    \\
   Bunch spacing (ns)                        & 25             & 25                   & 25            	    &  25       & 366 (350)           & 2$\times 10^{5}$	    \\
  Protons per bunch  ($10^{11}$)          & 1.7    & 2.2               & 2.5          	    & 2.5         & 2.2          & 2.2	    \\
  $\varepsilon_{p}$ ($\mu m$ rad)          & 3.7    & 2               & 2.5                         & 3.7	& 3.7          & 3.7	    \\
  IP $\beta^{*}_{p}$ (cm)                           & 10    & 7               & 10          	   & 10               & 10          & 10	    \\
 Pr. bunch length (mm)                               & 75.5    & 75.5              & 75.5              & 75.5	& 75.5          & 75.5	    \\  
Electrons per bunch ($10^{9} $)                & 0.045   & 0.045               & 0.045         &5.2 	& 17.4          & 10.0	    \\
 Electron current (mA)                                 & 0.3       & 0.3                     & 0.3          & 0.64	& 0.027          & 0.008	    \\
  $\varepsilon_{e}$ ($\mu m$ rad)              & 5           & 5                       & 5          & 5	          & 10                & 50	    \\
   IP $\beta^{*}_{e}$ (cm)                         & 120        & 44               & 44          	  &120        & 470          & 960	    \\
  El. bunch length (mm)                               & 0.21    & 0.21               & 0.21         &0.21   	& 0.225          & 0.020	    \\
Collision Frequency ($s^{-1}$)                   & 40$\times 10^{6}$     &40$\times 10^{6}$     &40$\times 10^{6}$         & 78$\times 10^{4}$    	& 9800          & 5000  \\
  CP to IP distance (cm)                             & 100    & 100               & 100          	&100            & 100          & 100	    \\
 Tot. Luminosity ($10^{30}cm^{-2}s^{-1}$)         & 6.2    & 10.0               & 11.7      & 14.0       	& 2.5          & 0.6	    \\
Lum. 0.9-1 $W_{max} $($10^{30}cm^{-2}s^{-1}$)   & 4.4    & 8.3               & 9.8       & 10.0      	& 0.8          & 0.2	    \\\hline
   \end{tabular}
%    \end{ruledtabular}
\end{table*}

\begin{figure}[htbp]
\centering
\includegraphics[width=0.55\columnwidth]{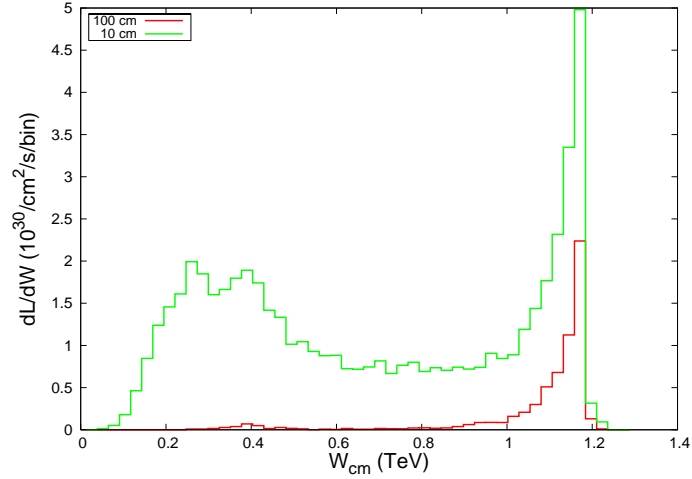}
\caption {Comparison of Luminosity spectra of LHeC ERL$\times$LHC for short and long CP to IP distance.}
\label{fig:lhecl10cm100cmspec}
\end{figure}

\begin{figure}[htbp]
\centering
\includegraphics[width=0.55\columnwidth]{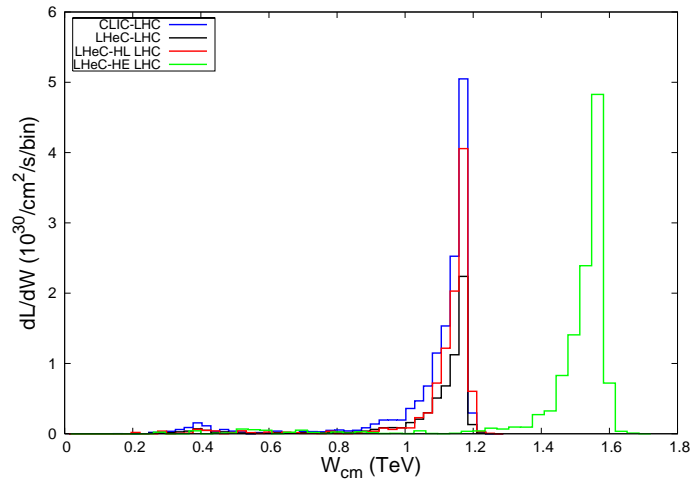}
\caption {Luminosity spectra for LHeC ERL$\times$LHC, LHeC ERL$\times$HL-LHC, LHeC ERL$\times$HE-LHC and CLIC$\times$LHC.}
\label{fig:lheclumispec}
\end{figure}

\begin{figure}[htbp]
\centering
\includegraphics[width=0.55\columnwidth]{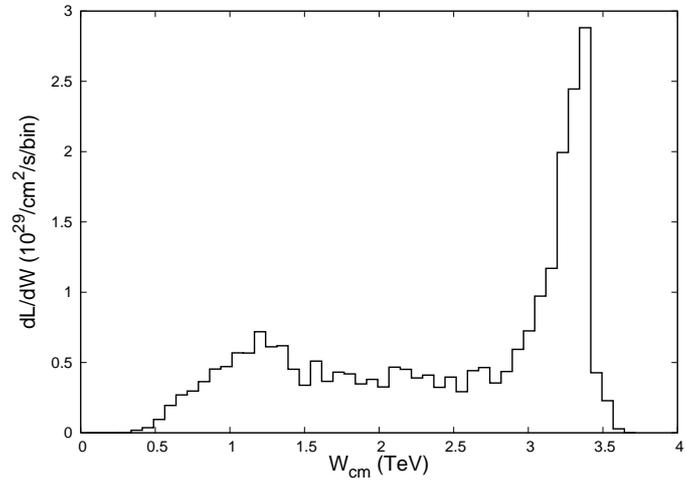}
\caption {Luminosity spectrum for ILC$\times$LHC.}
\label{fig:ILClumispec}
\end{figure}

\begin{figure}[htbp]
\centering
\includegraphics[width=0.55\columnwidth]{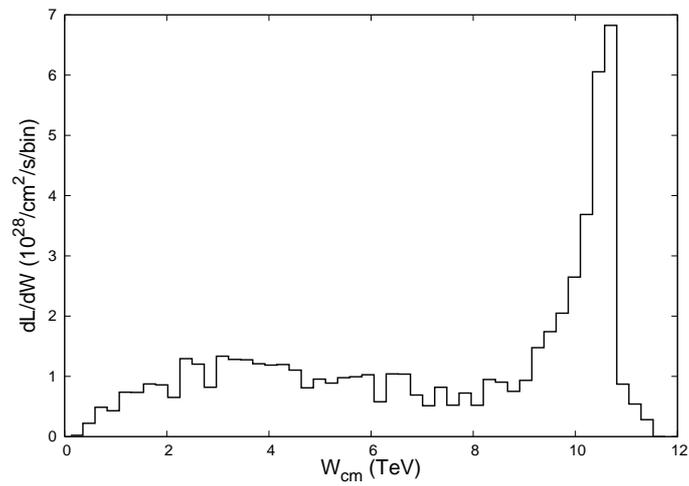}
\caption {Luminosity spectrum for PWFA-LC$\times$LHC.}
\label{fig:pwfalumispec}
\end{figure}

The luminosity distribution in terms of $W_{\gamma p}$ center of mass energy is 
\begin{equation}
\frac{dL_{\gamma p}}{dW_{\gamma p}}=\frac{W_{\gamma p}}{2 E_{p}}\frac{N_{\gamma} N_{p}f_{coll}}{2 \pi (\sigma_{e}^{2}+ \sigma_{p}^{2})}f(\frac{W_{\gamma p}^2}{4E_{p}})exp[-\frac{z^{2}\theta_{\gamma}(\frac{W_{\gamma p}^2}{4E_{p}})^2}{2 (\sigma_{e}^{2}+ \sigma_{p}^{2})}]
\label{eq:lumigp}
\end{equation}
where $E{p}$ is proton beam energy, $ f(\frac{W_{\gamma p}^2}{4E_{p}})$ signifies the differential Compton cross section, $N{\gamma}$ is the number of back scattered photons per pulse,  $N{p}$ is the protons per bunch,
$f_{col}$ is collision frequency, $\sigma_{e}$ anf $\sigma_{p}$ are transverse beam sizes electrons and protons, $\theta_{\gamma}(\frac{W_{\gamma p}^2}{4E_{p}})$ are angle of backscattered photons
and z is the distance between conversion and interaction points \cite{aksakal07}.  
The luminosity spectrum of gamma proton collider is strongly related with the distance between the conversion and interaction points. 
As it can be seen in Fig.~\ref{fig:lhecl10cm100cmspec} by increasing the distance the total luminosity is decreasing. However, the spectrum 
become more monochromatic. At high energies the effect of the distance on the luminosity is relatively low. Another advantage of the long distance
is to make relatively easier extraction of the spent electrons. Therefore the CP to IP distance is chosen 100 cm for all LHC based collider options. 
Luminosity spectra of $\gamma p$ colliders based on  LHeC and  LHC options are shown inFig.~\ref{fig:lheclumispec}. The luminosity spectra for  
ILC$\times$LHC and PWFA-LC$\times$LHC are presented in Figs.~\ref{fig:ILClumispec} and~\ref{fig:pwfalumispec}, respectively. 

\subsection {\label{sec:gfcc} FCC based $\gamma$p colliders}

\begin{figure}[htbp]
\centering
\includegraphics[width=0.55\columnwidth]{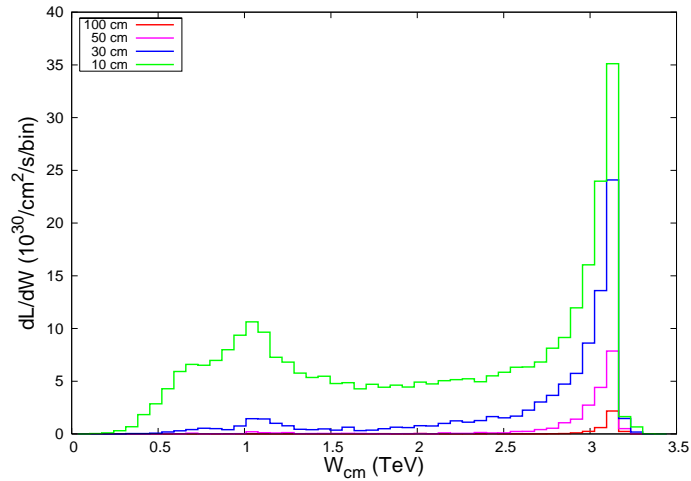}
\caption {Luminosity spectrum for LHeC ERL$\times$FCC for different CP to IP distances.}
\label{fig:lhecFCClumispec}
\end{figure}

\begin{figure}[htbp]
\centering
\includegraphics[width=0.55\columnwidth]{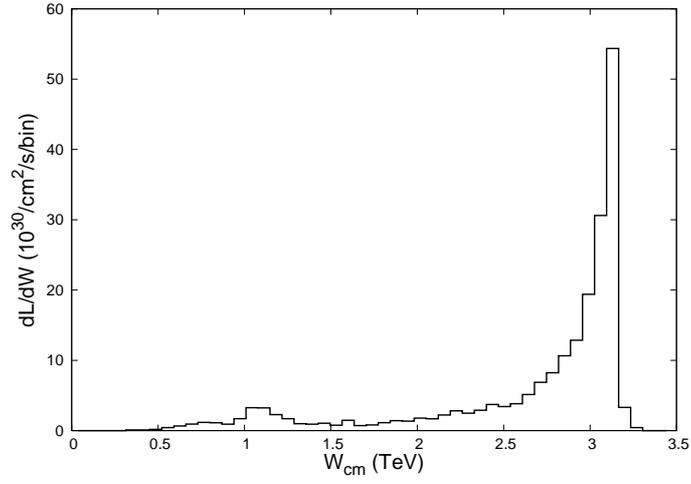}
\caption {Luminosity spectrum for CLIC$\times$FCC.}
\label{fig:CLICFCClumispec}
\end{figure}

\begin{figure}[htbp]
\centering
\includegraphics[width=0.55\columnwidth]{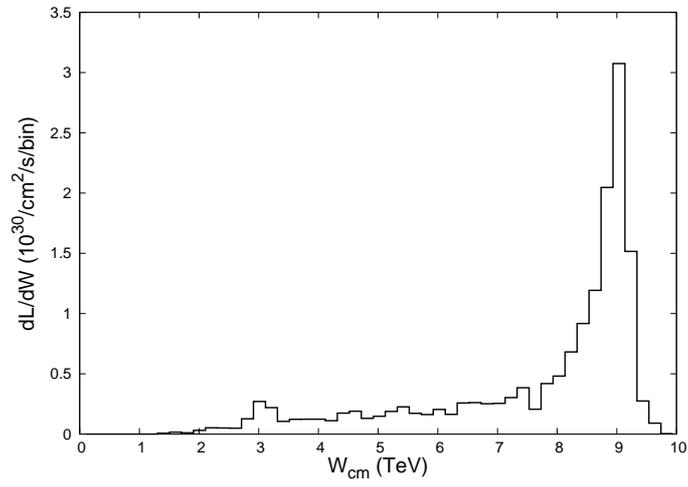}
\caption {Luminosity spectrum for ILC$\times$FCC.}
\label{fig:ILCFCClumispec}
\end{figure}

\begin{figure}[htbp]
\centering
\includegraphics[width=0.55\columnwidth]{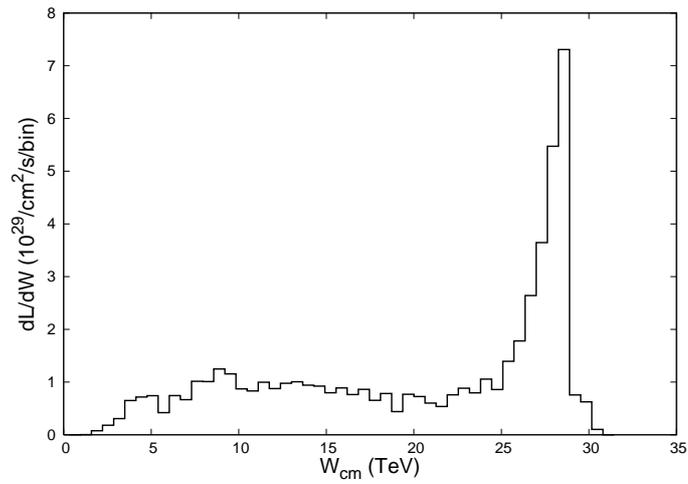}
\caption {Luminosity spectrum for PWFA-LC$\times$FCC.}
\label{fig:PWFCFCClumispec}
\end{figure}

FCC is the future project, which includes pp collider with 100 TeV center-of-mass energy, supported by European Union within the Horizon 2020 Framework Program for Research and Innovation \cite{benedict15}. Proposed linacs and FCC beam parameters for
 $\gamma p$ colliders based on FCC are given in Table~\ref{efccparam}. The bunch spacing of ILC is greater than FCC's bunch spacing. Therefore, most of the proton bunches would not interact with ILC's electrons. However, number of protons per bunch can be increased by decreasing the number of proton bunches. Upgraded parameters are shown in the table in parenthesis. Same upgrade also applied for PWFA-LC$\times$FCC.  Luminosity spectra of proposed  LHC based
$\gamma p$ colliders are shown in Figs.~\ref{fig:lhecFCClumispec}, ~\ref{fig:CLICFCClumispec}, ~\ref{fig:ILCFCClumispec} and~\ref{fig:PWFCFCClumispec}. 

\begin{table*}[htbp]
    \caption{Proposed accelerator parameters for different e-FCC based $\gamma$p colliders. }\label{efccparam}
\smallskip
\centering
\small\addtolength{\tabcolsep}{-4pt}
    \begin{tabular}{lcccc}\hline
    Parameters                          & FCC-eh	    & CLIC-FCC  & ILC-FCC  & PWFA-FCC \\\hline
  $E_{p}$  (TeV)            & 50          & 50              &50       & 50  \\
    $E_{e}$  (GeV)         & 60          &60               & 500        & 5000   \\
   Max. C.M.E (TeV)                          & 3.14    &3.14    & 9               & 28.7   \\
   Bunch spacing (ns)                        & 25       &25         & 366 (350)           & 2$\times 10^{5}$		    \\
  Protons per bunch  ($10^{11}$)          & 1.0   & 1         & 1.0 (2.2)               & 1.0 (2.2)  \\
  $\varepsilon_{p}$ ($\mu m$ rad)          & 2.2    &2.2       & 2.2               & 2.2  \\
  IP $\beta^{*}_{p}$ (cm)                           & 15    & 15        & 10               & 10      \\
 Pr. bunch length (mm)                         & 75.5   &75.5 & 75.5              & 75.5        \\
  Electrons per bunch ($10^{9} $)                & 0.045   & 5.2  & 17.4      & 10    \\
 Electron current (mA)                                 & 0.3   &0.64    & 0.027        &0.008      \\
  $\varepsilon_{e}$ ($\mu m$ rad)             & 5        &5          & 10               & 50      \\
   IP $\beta^{*}_{e}$ (cm)                      & 14        &14           & 40               & 80   \\
  El. bunch length (mm)                         & 0.210  & 0.21  & 0.225        & 0.020  	    \\
Collision Frequency ($s^{-1}$)          &40$\times 10^{6}$         & 78$\times 10^{4}$    & 9800          & 5000  \\
  CP to IP distance (cm)                         & 30   & 30       & 100               & 100     \\
 Tot. Luminosity ($10^{30}cm^{-2}s^{-1}$)         & 91.0 & 200   & 16      & 5.2   \\
Lumi. 0.9-1$W_{max}$($10^{30}cm^{-2}s^{-1}$)         & 50  & 114  & 8.3      & 1.9	    \\\hline
   \end{tabular}
%    \end{ruledtabular}
\end{table*}

\section{\label{sec:physics} Physics at $\gamma$p colliders}

Analyses performed for UNK+VLEPP \cite{alekhin91}, HERA+LC \cite{aydin96}, THERA  \cite{thera} and LHeC \cite{lhec11} have shown superiority of $\gamma p$ colliders compared 
with corresponding ep colliders for a lot of SM and BSM phenomena (small x gluon, anomalous interactions of t quark, q* and so on). 
Similar studies should be performed for FCC based  $\gamma p$ colliders. Below we list several examples of physics phenomena where  $\gamma p$ colliders 
have a huge potential. 

Concerning BSM physics, polarization of high energy photon beam (for details see \cite{aksakal07, ciftci95}) will give an opportunity to determine Lorentz 
structure of $\gamma qq^{*}$, $\gamma qt$ ,$\gamma WW$ and $\gamma ZZ$ vertices. It is important that at  $\gamma p$ colliders we deal with pure photon 
interactions, whereas at ep colliders both $\gamma$ and Z contribute to corresponding processes and their interactions cannot be separated.

As for SM physics,  $\gamma p$ colliders will give opportunity to measure total cross-sections of interaction of real photons with matter at very high energies comparable with cosmic $\gamma$-rays. 
Then, investigation of the process $\gamma p\rightarrow bbX$ will give opportunity to clarify QCD basics: $Q^{2}\geq 4m_{b}^2\approx100 \, GeV^{2}$  means perturbative 
QCD while $x_{g}\approx4 (m_{b}^{2})\diagup \sqrt{s_{\gamma p} }$ (see Table~\ref{phys}) correspond to high density of gluons (saturation region).

\begin{table}[htbp]
    \caption{Characteristic $x_{g}$ values for pair production of c and b quarks at different $\gamma p$ colliders. }\label{phys}
\smallskip
\centering
\small\addtolength{\tabcolsep}{-2pt}
    \begin{tabular}{lcccccc}\hline
 Protons                         & \multicolumn{3}{c}{LHC}	  &  \multicolumn{3}{c}{FCC} \\\hline
 Electrons            & ERL                  & ILC                             & PWFA                                & ERL                             &ILC                  &PWFA  \\
  $\overline{c}c$                     & $10^{-5}$       & $10^{-6}$                    & $10^{-7}$             	& $10^{-6}$              & $10^{-7}$        &$10^{-8}$     \\
  $\overline{b}b$                   & $10^{-4}$       & $10^{-5}$                    & $10^{-6}$             	& $10^{-5}$              & $10^{-6}$        &$10^{-7}$     \\\hline
   \end{tabular}
%    \end{ruledtabular}
\end{table}

\section {Conclusion}

Lepton-hadron collider with $\sqrt{s_{ep}}$ of order of 1 TeV (multi-TeV) is necessary both to clarify fundamental aspects of the 
QCD part of the Standard Model and for adequate interpretation of experimental data from the LHC (FCC-hh). Furthermore, 
linac-ring type ep colliders will provide an opportunity to construct $\gamma p$ colliders with  $\sqrt{s_{\gamma p}} \approx  0.9 \sqrt{s_{ep}}$. 

In this paper, we have described the design parameters, including laser requirements, for γp colliders based on LHC and FCC. 
The effect of distance between conversion and interaction points has been analyzed: more distance means more “monochromaticity” but less luminosity. 

Certainly, multi-TeV scale $\gamma p$ collider  have a huge search potential for both SM and BSM physics and essentially enlarge capacity of basic ep collider. 
In order to clarify this potential, systematic study of different physics phenomena is needed. We cordially invite HEP community to start this study.

\section {Acknowledgment}
This work was supported by the Turkish Atomic Energy Agency with Grant No.~2015 TAEK (CERN) A5.H6.F2-13 and also, in part, by the European Commission
under the HORIZON2020 Integrating Activity
project ARIES, grant agreement 730871.\\

\bibliographystyle{elsarticle-num}
%\section*{References}
\bibliography{gp1}

\end{document}